\documentclass[a4paper,english,12pt]{article}

\usepackage[version=3]{mhchem} 
\usepackage{fontenc}           
\usepackage{hyperref}
\usepackage[latin1]{inputenc}
\usepackage{float}
\usepackage{color}
\usepackage{graphicx}
\usepackage{amssymb}
\usepackage{babel}
\usepackage{amsmath}
\usepackage{fancyhdr}
\usepackage{geometry}
\usepackage{comment}

\usepackage[affil-it]{authblk}

\author[1]{Kalun Bedingfield}
\author[1]{Ben Yuen}
\author[1]{Angela Demetriadou\thanks{a.demetriadou@bham.ac.uk}}
\affil[1]{School of Physics and Astronomy, University of Birmingham, Edgbaston, Birmingham, B15 2TT, United Kingdom}

\title{Subradiant entanglement in plasmonic nanocavities}

\begin{document}

\maketitle

\begin{abstract}
Plasmonic nanocavities are known for their extreme field enhancement and sub-wavelength light confinement in gaps of just a few nanometers. 
Pairing this with the ability to host quantum emitters, they form highly promising platforms to control or engineer quantum states at room temperature. 
Here, we use the lossy nature of plasmonic nanocavities to form sub-radiant entangled states between two or more quantum emitters, that persist for $\sim 100$ times longer than the plasmonic excitation.
We develop a theoretical description that directly links quantum variables to experimentally measurable quantities, such as the extinction cross-section, and unlike previous studies includes plasmonic excitations necessary to resonantly form subradiant states.
This work paves the way towards engineering quantum entangled states in ambient conditions with plasmonic nanocavities, for potential applications such as rapid quantum memories, quantum communications and sensors.

\end{abstract}

Plasmonic nanocavities are formed by two or more nano-metallic structures separated by a nanometer sized gap~\cite{Emboras2016,Sigle2015,Benz2015}, and have drawn significant attention in recent years due to their ability to reveal light-matter interactions in ambient conditions~\cite{Zengin2015,Chikkaraddy2016a,FelixBenz2016,Hoang2016,Huang2019,Zhang2017,Yang2020}. Such structures confine electromagnetic fields at extremely sub-wavelength volumes~\cite{Chikkaraddy2016a,FelixBenz2016,Zhang2017,Yang2020,Hoang2015} and promise to generate quantum phenomena, like quantum entanglement and non-classical light at room temperature.
Until now, experimental investigations have primarily focused on far-field measurements for the combined system  (i.e. scattering and extinction cross-sections)~\cite{Zengin2015,Chikkaraddy2016a,FelixBenz2016,Hoang2016,Huang2019,Zhang2017,Yang2020,Hoang2015,Baumberg2019,Kishida2022}, while theoretical studies have focused on quantum scattering processes and spontaneous emission~\cite{Franke2019,Ren2020,Schmidt2016,Neuman2019}, with quantum dynamics usually explored for arbitrary cavities, and typically in rapidly decaying systems~\cite{Shlesinger2018,Franke2019,Franke2020,Carlson2021}.

Quantum entanglement serves as the foundation of many current quantum technologies, typically implemented in high-finesse cavities or other low loss systems, and often require complex cryogenic setups. 
Plasmonic nanocavities, characterized by their non-Hermitian (i.e. lossy) nature, present a significant challenge to the coherent control of quantum states. 
However, it has been proposed that dissipation can be utilized for quantum state preparation and entanglement~\cite{Kraus2008,Verstraete2009,Wang2013}. Such dissipative state engineering has been realised in traditional quantum optical systems to produce entanglement in atomic ensembles~\cite{Krauter2011}, Bell states and squeezed states of trapped ions~\cite{Lin2013,Kienzler2015}, but it often involves complex experimental set-ups operating at cryogenic temperatures.

In this paper, we show that nanoplasmonic systems are ideal for forming quantum states in ambient conditions through dissipation, by demonstrating that subradiant states can be formed in plasmonic nanocavities.
We first integrate quantised plasmons into quantum optical models, paying particular attention to both the accurate parametrization of input and output coupling and an accurate description of a resonant plasmon mode. This allows us to directly connect quantum variables with experimentally measurable quantities, ensuring consistency in the classical limit. 
Using this method, we reveal sub-radiant entanglement emerging between two or more QEs located in a plasmonic nanocavity, that persists for $\sim 100$ times longer than the lifetime of Rabi oscillations that are limited by the plasmon decay.
Unlike previous studies that adiabatically elliminate the photonic excitation and therefore form subradiant states driven by virtual transitions~\cite{Plenio1999,Gonzalez2011,Shlesinger2018}, we include the excitation of the plasmonic mode, which is important near resonance. 
We show that it is the lossy nature of plasmonic nanocavities that aids the emergence of the subradiant states. 
Hence, our work is an important bridge between theoretical input-output modes in quantum optics~\cite{Gardiner1985}  and descriptions of quantum plasmonics, which paves the way towards engineering quantum states via dissipation in ambient condition with plasmonic devices.

\section{Quantum description of plasmonic nanocavities}
The open and lossy nature of plasmonic environments creates a non-Hermitian system and places significant theoretical impediments for their incorporation into cavity quantum-electrodynamic (cQED) descriptions, since they were developed for high-finesse cavities. 
Recently, several quantum descriptions have been proposed that aim to overcome the non-Hermitian nature of the problem~\cite{Li2016,Cuartero2018,Franke2019}.
Even though each method is a significant breakthrough, this is still a developing topic with each newly proposed method having significant merits, but also limitations. 
Two notable examples are the 
Weisskopf-Wigner theory~\cite{Li2016,Li2018,Cuartero2018,Cuartero2020,Cuartero2021,Medina2021} that in general uses the spectral density of a plasmonic system calculated classically to perform multi-Lorentzian spectral fits for the plasmonic modes, and the symmetrizing orthonormalization transformation of quasi-normal mode (QNM) operators that produces plasmonic operators that canonically commute~\cite{Franke2019}. 
Although both methods are probably the current state-of-the-art for the quantization of plasmons, the first method does not easily allow for the decomposition of plasmon modes (including both the near- and far-fields), while the latter mixes the plasmonic modes during the symmetrizing orthonormalization transformation, producing a new set of canonical plasmonic operators that do not correspond to the initial QNM modes.

Here, we therefore introduce a general open quantum system formalism, where each plasmonic mode's resonant frequency, loss and coupling strength with a QE is obtained from classical QNM modes, and focus on how energy couples in and out of the system to obtain experimentally measurable quantities.
This formalism can be applied to other quantum descriptions, including~\cite{Franke2019,Medina2021}.
We start by decomposing the classical photonic response of the plasmonic system into quasi-normal modes (QNMs), using the QNMEig methodology~\cite{Yan2018,QNMEig}. 
This method employs an auxiliary-field eigenvalue approach to account for the dispersive nature of the metals in a linear manner. 
The eigenvectors provide the near-field of each plasmon mode with complex eigenfrequencies~\cite{Kristensen2015a,Bedingfield2023}  $\tilde{\omega}_{cav} = \omega_{cav} - i\kappa_{out}$, where $\omega_{cav}$ is the cavity resonant frequency and $\kappa_{out}$ the total decay rate  (radiative and Ohmic losses) of each plasmonic mode. 
Here, we consider the Nano-Particle on Mirror (NPoM) plasmonic nanocavity as shown in Figure 1(a)-(b). 
\begin{figure}[!b]
\includegraphics[width=\linewidth]{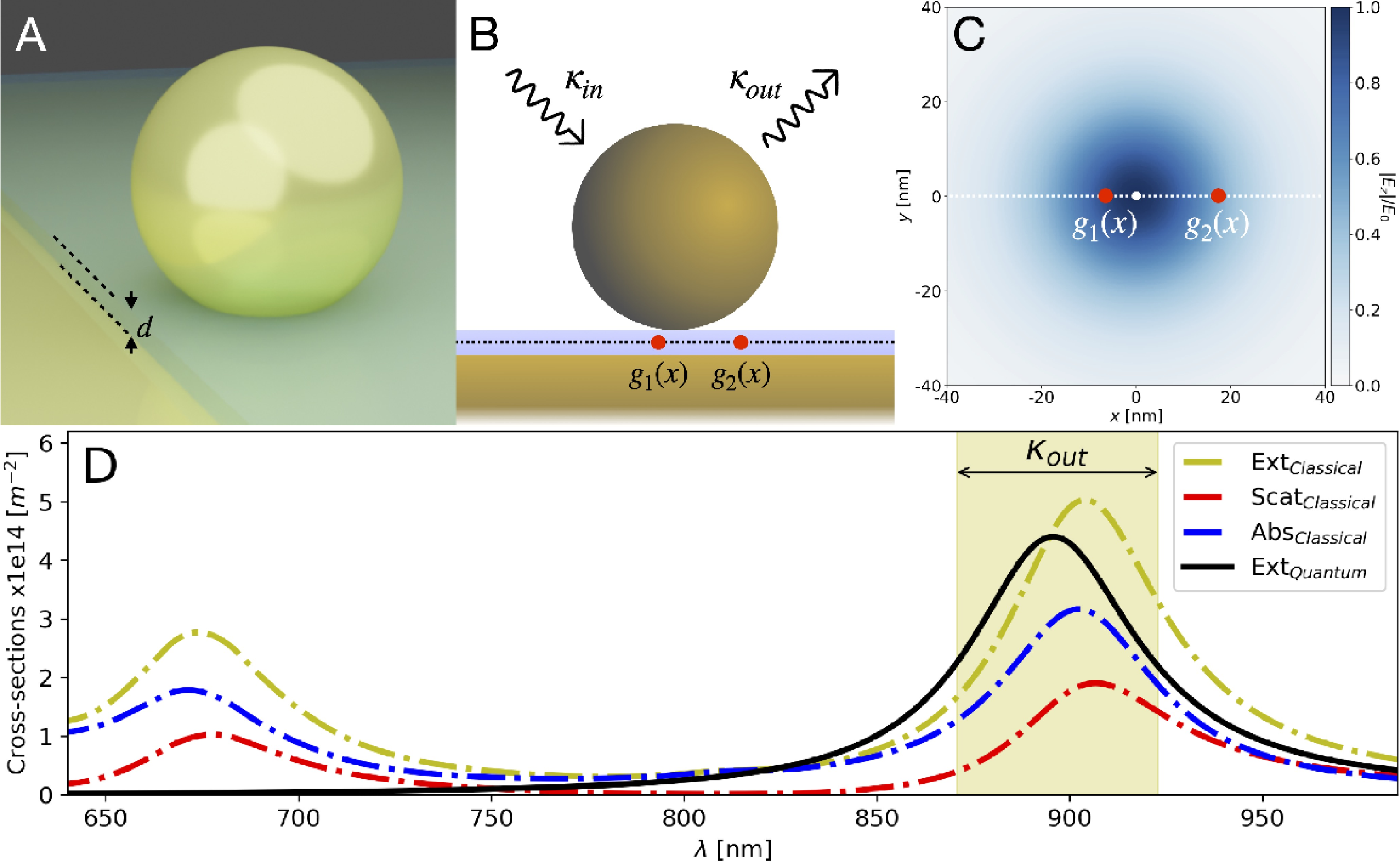}
\caption{\label{fig:n1}
(a) NPoM geometry for a $40$nm radius spherical gold nanoparticle assembled on a flat gold mirror, and separated by a spacer of refractive index $n=2.5$ and thickness $d=1$nm. 
(b) NPoM with two QEs in the nanocavity, and showing the energy coupling in ($\kappa_{in}$) and out ($\kappa_{out}$) of the system. 
(c) The normalised mode profile of the $(1,0)$ plasmonic mode on a plane through the centre of the cavity, with the white and red dots indicating the QE positions. 
(d) The classical absorption, scattering and extinction cross-sections with the quantum extinction cross-section for a single plasmonic mode model.}
\end{figure}

We divide the Hamiltonian $\mathcal{H}=\mathcal{H}_{sys}+\mathcal{H}_{res}$ into two components: the system ($\mathcal{H}_{sys}$) and reservoir ($\mathcal{H}_{res}$) Hamiltonians, which we discuss separately. The system Hamiltonian describes the plasmonic modes obtained from classical QNM calculations~\cite{Yan2018} ($\mathcal{H}_{cav}$), the multiple QEs as two-level systems ($\mathcal{H}_{QEs}$), the interaction between the plasmon modes and the QEs ($\mathcal{H}_{int}$) and a driving external field coupled to the cavity ($\mathcal{H}_{ext}$), such as a pump, that here represents an external illumination of the system. The system Hamiltonian is given by:
\begin{equation}
\mathcal{H}_{sys} =  \underbrace{\Delta_{cav} a^\dagger a}_{\mathcal{H}_{cav}} + \underbrace{\sum_j^N  \frac{1}{2} \Delta_d^{(j)} \sigma^{(j)}_z}_{\mathcal{H}_{QEs}} + \underbrace{\sum_j^N g(\mathbf{r}^{(j)}) \left( a^\dagger \sigma^{(j)}_- +  a \sigma^{(j)}_+ \right)}_{\mathcal{H}_{int}}  +\underbrace{ i \sqrt{\kappa_{in}} \left( \alpha^* a - \alpha a^\dagger \right)}_{\mathcal{H}_{ext}} \label{eq:HSys} 
\end{equation}
where $N$ is the number of QEs placed in the nanocavity, $\Delta_{cav} = \omega_{cav} - \omega_p$ and  $\Delta_d^{(j)} = \omega_{QE}^{(j)} - \omega_p$  are the detunings of the cavity resonance ($\omega_{cav}$) and each QE resonant frequency ($\omega_{QE}^{(j)}$)  from the external field's frequency ($\omega_p$). The $\left\{a,a^\dagger\right\} $ are the  bosonic operators that create and annihilate the plasmon mode, and $\sigma_{z}^{(j)}$ is the Pauli $z$-operator describing the $j$-th two-level QE (see Supp. Info.). Note that to ease the interpretation of the system's quantum dynamics and the correlations between multiple QEs residing within the plasmonic nanocavity, we design a gold NPoM nanocavity for which it is safe to assume that only one plasmon mode couples to the QEs. 
The interaction of the plasmonic mode with the $j$-th QE at position $\mathbf{r}^{(j)}$ is given by $\left( a^\dagger \sigma^{(j)}_- +  a \sigma^{(j)}_+ \right)$ with the coupling strength measured by:
\begin{equation}
g(\mathbf{r}^{(j)}) = \sqrt{\frac{\hbar \omega_{cav}}{2 \varepsilon_0 \varepsilon V}} \ \mathbf{d}^{(j)} \cdot \mathbf{u}(\mathbf{r}^{(j)})  \label{eq:couplstrength}
\end{equation}
where  $\mathbf{u}(\mathbf{r}^{(j)})$ is the normalized electric field vector of the QNM mode, which makes the coupling strength $g(\mathbf{r}^{(j)})$  dependent on the $j$-th QE's location $\mathbf{r}^{(j)}$  within the plasmonic nanocavity. The mode volume is obtained from the QNM calculations as $V = \frac{\int d \mathbf{r} |\mathbf{u} (\mathbf{r}) |^2}{\text{Max}[|\mathbf{u} (\mathbf{r}) |^2]}$  (see Supp. Info.), $\varepsilon$ is the electric permittivity of the material hosting the QEs and $\mathbf{d}^{(j)}$ is the dipole moment of the $j$-th QE.
The last term ($i \sqrt{\kappa_{in}} \left( \alpha^* a - \alpha a^\dagger \right)$ ) of equation~\eqref{eq:HSys} describes an external coherent monochromatic source driving the system, where $\sqrt{\kappa_{in}}$ is the rate that energy couples into the system and $\alpha$ is the amplitude of the coherent state, defined by the source's photon flux $c|\alpha|^2$ (see Supp. Info. for more discussion).

While the system Hamiltonian describes  the excitation of the system and the interaction between QEs and the plasmon mode, it contains no description of losses.
Nanoplasmonic devices hosting QEs in general have two loss channels: (i) the plasmon modes lose energy by both radiating to the far-field and via Ohmic losses, and (ii) the decoherences of QEs due to the (ro)-vibrational energy states of the molecule. 
We represent the plasmonic loss and QE decoherences with two separate reservoirs as:
\begin{eqnarray}
\mathcal{H}_{res}  = & \sum^N_j \int^\infty_0 d\omega_{vib} \ \sigma^{(j)}_z \kappa_{vib} (\omega_{vib}) \left[ b_{vib}^\dagger e^{i \omega_{vib} t} + b_{vib} e^{-i \omega_{vib} t} \right] \nonumber \\
& - i \sum_\lambda \int^\infty_0 d\omega \kappa_{res} (\omega) \left[ b_\lambda (\omega) a^\dagger e^{i \Delta(\omega) t} - b^\dagger_\lambda (\omega) a e^{-i \Delta(\omega) t} \right] \label{eq:HRes}
\end{eqnarray} 
where the first term describes the coupling of the QEs to a phonon bath reservoir, with phonon frequencies $\omega_{vib}$ and strength 
$\kappa_{vib}(\omega_{vib})$, and $\left\{b_{vib},b_{vib}^\dagger\right\}$ the phonon reservoir operators. 
The second term of equation~\eqref{eq:HRes}  accounts for the coupling of the plasmonic mode to a background plasmon reservoir, with photon frequencies $\omega$ and coupling strength $\kappa_{res}(\omega)$, to account for the energy lost both radiatively and via Ohmic losses. Note that the detuning $\Delta(\omega)=\omega-\omega_p$ is the detuning of frequency $\omega$ with the incident external field frequency ($\omega_{p}$), and $\left\{b_{\lambda}(\omega), b_{\lambda}^\dagger(\omega)\right\}$ are the plasmon reservoir creation/annihilation operators.

The time evolution of the open quantum system with the plasmon loss and QE decoherences is described by the Lindblad master equation as: 
\begin{align}
\frac{d}{dt} \rho  =& -i \left[ \mathcal{H} _{sys}, \rho \right] + 2 \kappa_{out} a \rho a^\dagger - \kappa_{out} \left( a^\dagger a \rho + \rho a^\dagger a \right) \nonumber \\
& + \sum^N_j \left[ \kappa_{vib} \sigma^{(j)}_z \rho \sigma^{(j)}_z - \frac{\kappa_{vib}}{2} \left( \sigma^{(j) 2}_z \rho + \rho \sigma^{(j) 2}_z \right) \right]  \label{eq:LME}
\end{align}
where the density operator $\rho$ describes the statistically mixed state of the system. The expectation value of any operator of the system can be obtained via $\langle O \rangle = \text{Tr}[O\rho]$, for example the expected plasmon excitation number $\left\langle a^\dagger a \right\rangle$.

\subsection{Quantum Extinction Cross-Section}
To obtain experimental quantities, such as the extinction cross-section, from the above cQED description (or any other quantum description), one needs a formalism on how energy couples in and out of the system. 
Here we represent a constant external illumination with a monochromatic plane wave by a single mode, $\omega_p$, of the plasmon reservoir that is coherently excited. The field operators $b_{\lambda}(\omega_p)$ are replaced by complex amplitudes $\alpha$ under the mean-field approximation, and are directly related to the incident photon (see Supp. Info.).
The rates with which energy is coupled into ($\kappa_{in}$) and lost ($\kappa_{out}$) from the system is schematically shown in Figure~\ref{fig:n1}b and can be used to find the quantum extinction cross-section.

Classically, the extinction cross-section is defined as the total power lost by the system ($P_{out}$) relative to the incident intensity ($S$) of the plane wave. When considering their quantum analogues in steady-state, they respectively become: $P_{out} = 2\kappa_{out} \hbar \omega \langle a^\dagger a \rangle$ and $S = c_{0} \hbar \omega |\alpha|^2$. The $\langle a^\dagger a \rangle = \frac{\kappa_{in} |\alpha|^2}{\kappa_{out}^2 + \Delta_{cav}^2}$, and is proportional to $\langle a \rangle$ and $\langle a^\dagger \rangle$ (see Supp. Info.), which leads to the quantum extinction cross-section:
\begin{align}
\langle \sigma_{ext}^q \rangle_{no QEs} = \frac{2\kappa_{in} \kappa_{out}}{c_{0}} \frac{1}{\kappa_{out}^2 + \Delta_{cav}^2} \ , \label{eq:QExtCS}
\end{align}
where $c_{0}$ is the speed of light constant. 
We define the in-coupling coefficient as: $\kappa_{in} = \kappa_{out} c_{0} \sigma_{ext}/2$, which ensures that the quantum, $\langle \sigma_{ext}^q \rangle_{no QEs} $ and classical, $\sigma_{ext}^c$, extinction cross-sections are equal $\langle \sigma_{ext}^q \rangle_{no QEs} = \sigma_{ext}^c$ when the external illumination is on resonance with the QNM mode (i.e. $\omega_{cav}=\omega_{p}$ and $\Delta_{cav} = 0$).

Now, we apply this theoretical description to a specific plasmonic system. Throughout this article, we consider a NPoM geometry (see figure~\ref{fig:n1}a-b) made of a spherical, gold nanoparticle of radius $40 \, nm$ assembled $1\,nm$ above a flat gold mirror, separated by a molecular monolayer with refractive index $n=2.5$, which is consistent with recent experimental set-ups~\cite{Simoncelli2016,Chikkaraddy2018}. 
This plasmonic nanocavity has the first order $(1,0)$ quasi-normal mode resonance at $\tilde{\omega}_{cav} = (338.9-i9.8)THz$ (i.e. $\lambda_{cav} \sim 900\, nm$). The normalised electric field eigenvector $\mathbf{u}(\mathbf{r})$ for the $(1,0)$ mode is shown on a slice through the centre of the cavity in Figure~\ref{fig:n1}c, and exhibits a large central extrema, with the mode volume being $\text{Re[}V\text{]}=60.01$nm$^3$. 

In Figure~\ref{fig:n1}d, we plot the classical extinction ($\sigma_{ext}$), scattering ($\sigma_{scat}$)  and  absorption ($\sigma_{abs}$) cross-sections that show the lowest order mode to be spectrally isolated, and the $\kappa_{out}$ obtained from the QNM calculations corresponding to the full-width half-maximum of $\sigma_{ext}$.
Therefore we can safely assume that only one plasmonic mode interacts with the QEs for this specific plasmonic nanocavity. 
The quantum extinction cross-section is in good agreement with its classical counterpart, and since we considered only one plasmonic mode for the cQED description, it only shows one peak. 
Note that  the values of the quantum and classical $\sigma_{ext}$ match exactly on resonance with the cavity (i.e. the $(1,0)$ QNM mode: $\lambda_{(1,0)} \sim 900 \,$nm). 
The spectral shifts seen in  Figure~\ref{fig:n1}D, are due to the broadband contributions of higher order modes for far-field quantities such as the classical $\sigma_{ext}$, and have been previously reported and explained~\cite{Lombardi2016,Alonso2013}.
This new driven formalism is applicable to other quantum descriptions and very flexible, allowing us to initialise the system in any particular state, unlike other methods~\cite{Li2016}.

\section{Two-Emitter Quantum Dynamics}
\label{sec:2QEDyn}
We now consider the quantum dynamics of QEs in this cavity, which we take henceforth as Cy5 molecules with $d = 10.1$D and $\kappa_{vib}=25$meV~\cite{Chikkaraddy2018}. Before discussing the details of the two-molecule dynamics, we briefly summarise the single molecule behaviour when a plasmon is initially excited and the molecule is located in the centre of the cavity ($x=0$) and starts in its ground state. In this case Rabi oscillations occur at $\omega_{Rabi}=\sqrt{4g^2-\kappa^2_{out}}$. The molecular dephasing at rate $\kappa_{vib}$ causes these oscillations to 'blur' out over time (see Supp. Info.). Introducing a second emitter at the same location and initialisation one finds the frequency of Rabi oscillations is increased to $\omega_{Rabi}=\sqrt{8g^2-\kappa^2_{out}}$ due to the $\sqrt N = \sqrt 2$ enhancement of the collective coupling strength, $\sqrt 2 g$. The populations of the two emitters also decay faster due to the combined dephasing of both emitters.

\begin{figure}
\includegraphics[width=\linewidth]{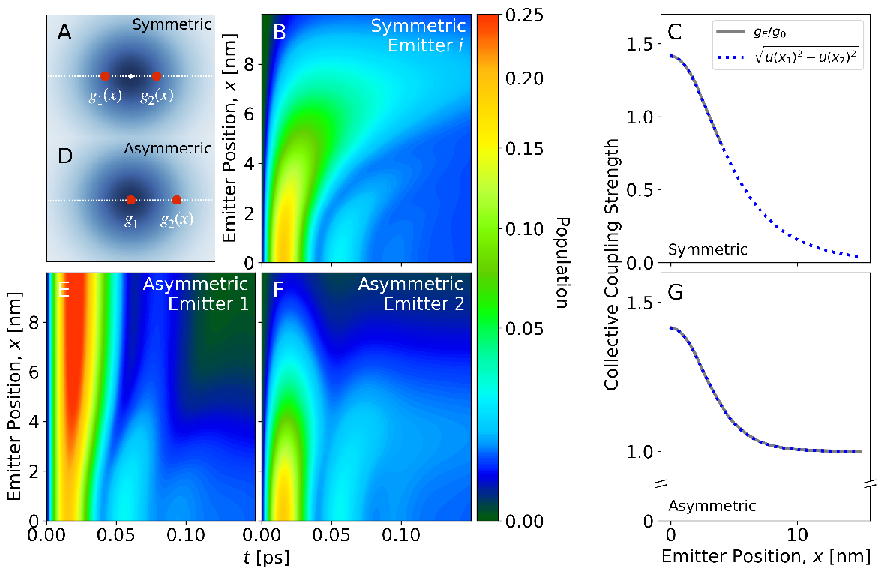}
\caption{\label{fig:n2}
(A) The symmetric QE placement in a NPoM nanocavity, (B) the population of their excited state, when the system is initialised in the excited field state and (C) the effective coupling strength of the system. (D) Assymetric QE placement in the nanocavity, with the (E) central and (F) displaced QE excited state population, when the system is initialised in the excited field state and (G) the system's effective coupling strength. 
}
\end{figure}
Realistically, two Cy5 molecules cannot be placed at the exact same position within the nanocavity, and their coupling strength $g(\mathbf r)$ of equation~\ref{eq:couplstrength} is position dependent, according to the  QNM mode's eigenvector shown in Figure~\ref{fig:n1}c.
Due to the cylindrical symmetry of the cavity and of the $(1,0)$ mode, we need to only consider QE positions along the $x$-axis---as indicated by the white dashed line in Figures~\ref{fig:n2}a and d. 
Figure~\ref{fig:n2}a diagrammatically shows a symmetric arrangement for the two QEs away from the centre of the cavity, which means both QEs continue to experience the same coupling strength, producing identical excited state populations for all positions $|x|=x_{2}=-x_{1}$. 
These are shown in Figure~\ref{fig:n2}b for $|x|$ up to $10$nm away from the nanocavity centre.
As the emitters move away from the cavity centre, they experience a rapid reduction of their coupling strength, and for $|x|> 6$nm the system transitions into the weak coupling regime. 
This can be quantified through the ratio of an effective coupling strength $g_E$ relative to the coupling of a single emitter at the centre of the cavity, $g_0$, which is shown in Figure~\ref{fig:n2}c. 
Due to the dependence of the coupling strength on the mode profile, the effective coupling strength ($g_{E}$) changes with the position of the emitters as: $g_{E}/g_{0} \propto \sqrt{u(x_1)^2 + u(x_2)^2}$, which becomes $\sqrt{2}u(x)$ for the symmetric placement of the QEs (since $u(x_1)= u(x_2)$).
If instead one QE always remains at the centre of the cavity, while the second QE takes various positions along the x-axis (i.e. asymmetric placement) as shown in Figure~\ref{fig:n2}d, then each QE is coupled to the plasmon mode with a different coupling strength. 
The population dynamics of the central emitter and displaced emitter now differ, and are respectively shown in Figure~\ref{fig:n2}e and f, showing that the central QE's Rabi oscillations slow down as the second emitter is placed further out from the nanocavity. 
Nevertheless, we find that the Rabi oscillations are still governed by the effective coupling strength $g_{E}/g_0\propto  \sqrt{u(x_1)^2 + u(x_2)^2}$ (see Figure~\ref{fig:n2}g). This now decreases more slowly as $\sqrt{1 + u(x_2)^2}$ compared to the symmetric case, and reduces to unity once the second QE is only weakly coupled to the plasmon, which occurs  at $x_2 \sim 10\, nm$ and the  system returns to a single emitter interaction.

\subsection{Hybrid states}
\label{sec:PumpDyn}
Experimentally the Rabi oscilaltions and effective coupling strength are obtained by the Rabi splitting of the system's hybrid states, measured in the far-field. 
Usually an external field illuminates the combined system, and the scattered light (or extinction energy) collected. 
Due to reciprocity, the energy loss of the system in steady state ($P = 2\kappa_{out} \hbar \omega \langle a^\dagger a \rangle$) is equivalent to the plasmon excitation in the system by the external illumination, which is described by: $P = - \sqrt{\kappa_{in}} ( \alpha \langle a^\dagger \rangle + \alpha^* \langle a \rangle)$. Normalizing this energy loss with the incident energy $S=c_0\hbar \omega |\alpha|^2$ leads to the quantum extincion cross-section for the combined system: $\langle \sigma_{ext}^q \rangle = - \sqrt{\kappa_{in}} ( \alpha \langle a^\dagger \rangle + \alpha^* \langle a \rangle)/c_{0}|\alpha|^2 $, in terms of the field operator expectation values. Without loss of generality, $\alpha$ can be taken to be real in our case, which reduces the quantum extinction cross-section to:
\begin{align}
\langle \sigma^q_{ext} \rangle = - 2 \sqrt{\kappa_{in}} \frac{Re\{\langle a \rangle \} }{ c_{0} \alpha}
\end{align}
where $\langle a \rangle$ describes the rate at which energy is removed from the excited plasmon state (i.e. extinction energy) and can be found by first solving the master equation for the density matrix in steady state, and then taking the trace with the field operator as: $\langle a \rangle = \text{Tr}[a \rho]$.

\begin{figure}
\includegraphics[width=\linewidth]{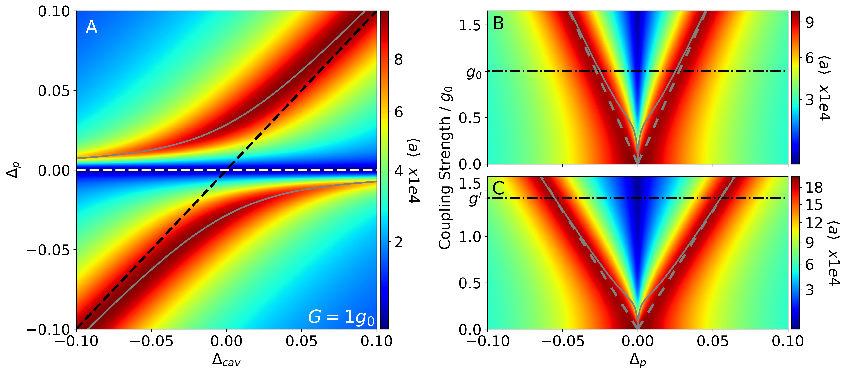}
\caption{\label{fig:n3}(A) The $\langle a \rangle$ of the driven single QE NPoM system as a function of both cavity ($\Delta_{cav} = \frac{\omega_{cav} - \omega_{em}}{\omega_{em}}$) and external source ($\Delta_p = \frac{\omega_p-\omega_{em}}{\omega_{em}}$) detunings. The black and white dashed lines respectively show $\Delta_{cav} = \Delta_{p}$ and $\Delta_{p} = 0$. Rabi splitting in terms of $\langle a \rangle$ as a function of both plasmon detuning and coupling strength, for (B) one and (C) two emitters at the centre of the cavity, where $g'=\sqrt{2}g_{0}$. Grey dashed lines trace the peaks, while the solid grey lines are solutions of equation~\ref{eq:hybrid_states}.}
\end{figure}

For this externally driven regime, the system which is initially in the vacuum state, is excited at a rate $\sqrt{\kappa_{in}} \alpha$---such that the average photon number in the system is $N_{ph}=10^{-6}$. In Figure~\ref{fig:n3}a, we plot $\langle a \rangle$ (which is proportional to $\langle \sigma_{ext}^q \rangle$) for a single emitter at the centre of the nanocavity, with both the external source ($\Delta_p$) and cavity ($\Delta_{cav}$)  resonances detuned relative to the QE resonance. 
This exhibits the signature avoided crossing to form two hybrid states, characteristic of strong coupling~\cite{Eck1967,Torma2014}. The black and white dashed lines show $\Delta_{cav} = \Delta_{p}$ and $\Delta_{p} = 0$ respectively.
Many experimental analyses of strongly coupled systems measure these hybrid states, with other theoretical descriptions also predicting them\cite{Demetriadou2017,Cuartero2020,Wu2021}. 
It has been recently shown~\cite{Wu2021} that one can consider a complex mode volume that accounts for the radiative nature of plasmons, to obtain the complex frequency of the strong coupling hybrid states as~\cite{Wu2021,Andreani1999,Santhosh2016,Park2019}
\begin{align}
\tilde{\omega}_{\pm} = \frac{\tilde{\omega}_e + \tilde{\omega}_{cav}}{2} \pm \sqrt{g^2 [1 - iR] + \left( \frac{\tilde{\omega}_e - \tilde{\omega}_{cav}}{2} \right)^2} \label{eq:hybrid_states}
\end{align}
where $R=\text{Im[}V\text{]/Re[}V\text{]}$ is the ratio of the imaginary and real components of the mode volume~\cite{Lalanne2018}, and the imaginary components of $\tilde{\omega}_e$ and $\tilde{\omega}_{cav}$ are respectively $\kappa_{vib}$ and $\kappa_{out}$. For systems with a large quality factor, $Q$, the negligibly small $\text{Im[}V\text{]}$ makes it apparent that the standard description is sufficient. For low Q systems though, such as our plasmonic nanocavity, the complex nature of the mode volume has the ability to significantly alter the observed dynamics. The solutions for $\tilde{\omega}_{\pm}$ are plotted with grey lines in Figure~\ref{fig:n3}a, demonstrating their strong agreement with our non-Hermitian cQED description for light-matter interactions.

Now, we obtain the strong coupling hybrid states for various coupling strengths with respect to the external field detuning from the QE $\Delta_{p}$ (see Figure~\ref{fig:n3}b). 
For an empty cavity ($g=0$), a singular resonance is observed which is attributed to the absorption and re-emission of the energy collected by the plasmonic mode alone. 
Increasing the coupling between the QE and the plasmon, one quickly sees their hybridisation and emergence of the characteristic Rabi splitting, even for small coupling strengths. 
For small coupling strengths the strong coupling condition $g^2 > \kappa_{out} \kappa_{vib}$ is still valid, despite the QE population dynamics showing a relatively small number of Rabi oscillations that dissipate quickly.
However, to acquire a more intuitive picture of the occurring interaction, it is more appropriate to separate this condition into two: one for the plasmon $g > \kappa_{out}$ and one for the QE $g > \kappa_{vib}$. 
For example, a single emitter at the centre of our NPoM cavity (i.e. at $g_{0}$) has: $g_{0}^2 / \kappa_{out} \kappa_{vib} = 4.64$, placing the system well into the strong coupling regime. However, $g_{0} / \kappa_{vib} = 9.74$, which allows the Rabi splitting to occur, while $g_{0} / \kappa_{out} = 0.48$, which leads to a large radiative emission and rapidly decaying Rabi oscillations. 
To lose the Rabi splitting and the hybrid states for the overall system, the coupling strength needs to reduce enough for the ratio $g / \kappa_{vib} <1$, which is in agreement with both our description and 
equation~(\ref{eq:hybrid_states})~\cite{Wu2021} (grey lines) shown in Figure~\ref{fig:n3}b. 
For multiple QEs in the cavity, the coupling strength scales with $\sqrt{N}$, as expected, with  Figure~\ref{fig:n3}(c)) showing the Rabi splitting for varying coupling strengths of two emitters in a plasmonic nanocavity obtained from our description and also in agreement with equation~(\ref{eq:hybrid_states}).

\section{Sub-radiant Entanglement}
\label{sec:2QESPE}

\begin{figure} 
\begin{center}
\includegraphics[width=\linewidth]{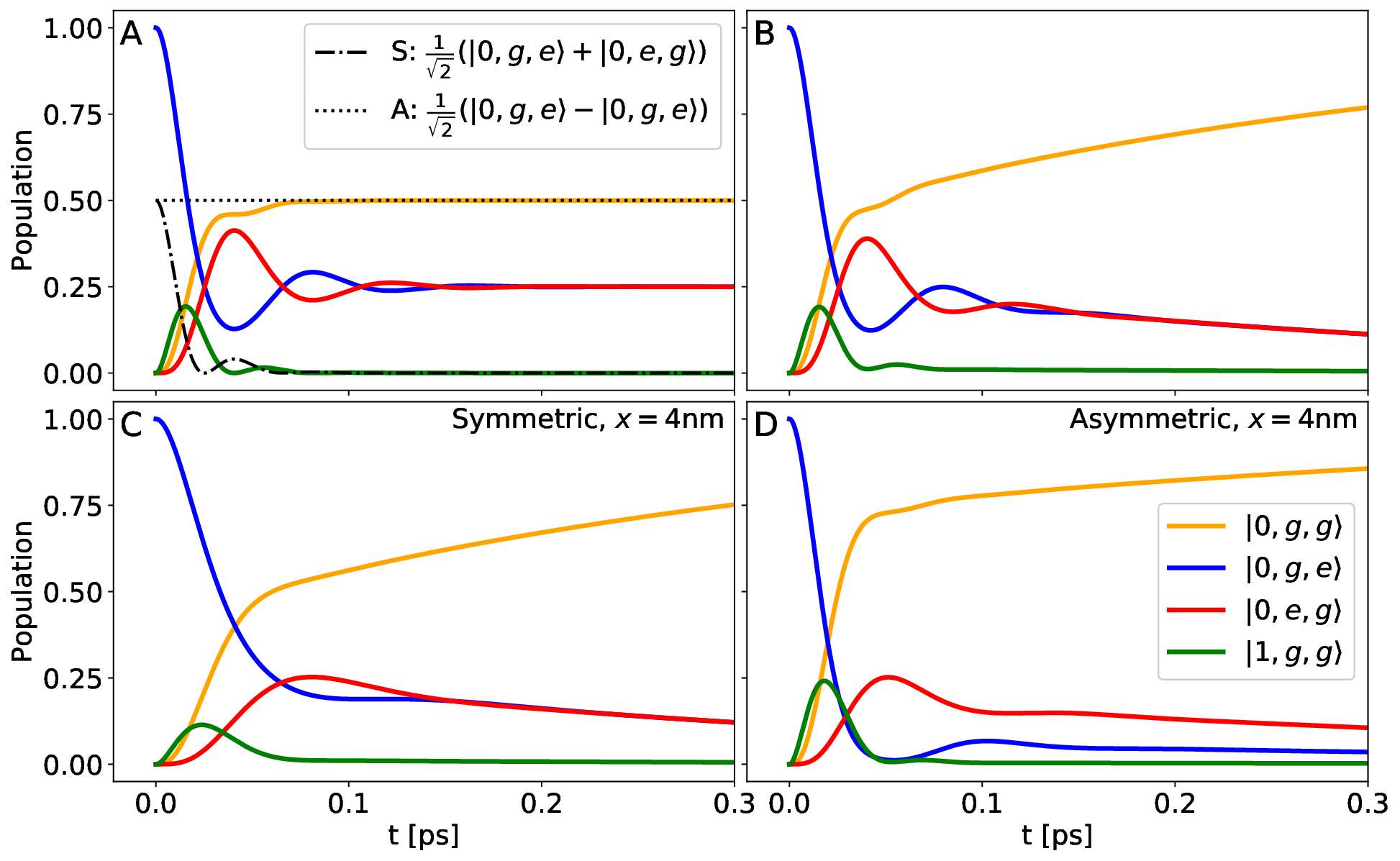}
\end{center}
\caption{\label{fig:n4}The population dynamics for a system of two QEs at the centre of a nanocavity (A) without ($\kappa_{vib}=0$) and (B) with ($\kappa_{vib}=25$meV) the QE decoherences.
S and A respectively describe the super- and sub-radiant superposition states. The population dynamics for two QE with decoherences $\kappa_{vib}=25$meV, when they are placed in a nanocavity (C) symmetically ($x_2=-x_1=4$nm) and (D) asymmetrically ($x_1=0$nm and $x_2=4$nm). 
} 
\end{figure}
We now take advantage of the flexibility of our cQED description to consider no external illumination driving the system, by setting the input field $\alpha$ to zero, and initialize the system with only one QE in the excited state. 
To better demonstrate the unique dynamics presented by this preparation, we initially ignore the inherent molecular linewidths of the emitters ($\kappa_{vib}=0$), and assume both emitters are placed at the centre of the cavity. 
Figure~\ref{fig:n4}A shows that the primary energy exchange and Rabi oscillations are now observed between the two emitters---with twice as much energy passed to the initially non-excited emitter than to the plasmonic mode. 
Due to the asymmetric initialisation, the emitters no longer interact with the field symmetrically: they instead oscillate out of phase with each other as energy is passed between them. The key observation here is that the emitters exchange energy until they reach an equal, non-zero steady state value. 
This long lived state originates from the symmetric and anti-symmetric superpositions formed from the states of the two emitters: 
\begin{align}
|S\rangle & = \frac{1}{\sqrt{2}} \left( |0,e,g\rangle + |0,g,e\rangle \right) \\
|A\rangle & = \frac{1}{\sqrt{2}} \left( |0,e,g\rangle - |0,g,e\rangle \right) \ ,
\end{align}
which respectively describe the super- and sub-radiant states, as shown in Figure~\ref{fig:n4}A with dot-dashed and dotted lines. 
Although the super-radiant state rapidly decays to zero, the sub-radiant state reaches a steady state solution. 
This occurs due to the asymmetry of the state, with the interactions of the $|1,g,g\rangle$ state with the components of $|A\rangle$ ($|0,e,g\rangle$ and $|0,g,e\rangle$) perfectly cancelling out. 
In fact, these states are formed via the high dissipation of the plasmonic mode, with the plasmon providing a conduit to form the entangled state. 
Introducing the molecular decoherence rates to this calculation (Figure~\ref{fig:n4}B), we find that these states are still formed, but now decay non-radiatively via $\kappa_{vib}=25$meV and eventually the system returns to its ground state. However, they live for $\sim 100$ times longer than the plasmonic excitation, creating semi-persistent sub-radiant entangled states in lossy plasmonic nanocavities.

For a more realistic arrangement of the emitters, Figure~\ref{fig:n4}C shows the quantum dynamics for a symmetric placement of QEs at $|x|=x_{2}=-x_{1}=4$nm.
The equal coupling strength leads to less oscillations between the two QEs, but eventually the same long lived entangled states emerges, with the same population as with both QEs at the centre of the nanocavity (Supp. Info.). 
In contrast, the asymmetric QE placement  (i.e. $x_{1}=0$nm and $x_{2}=4$nm) leads to unequal coupling strengths with the plasmon, and different rates for the energy exchange between them, which distorts the sub-radiant entangled states.
Interestingly, we conclude that the coherence between emitters and the formation of the sub-radiant entangled states occurs via the plasmonic dissipation and is only dependent on their relative coupling strength with each other.

\section{Multi-Emitter systems}
\label{sec:8QEDyn}
\begin{figure} 
\includegraphics[width=\linewidth]{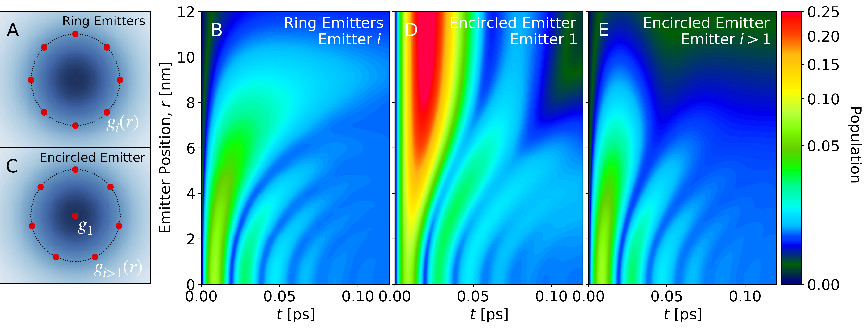}
\caption{\label{fig:n5}
(A) Eight QEs placed in a nanocavity on a ring arrangement, and (B) the population dynamics of their excited state, when the system is initialised in the excited field state. (C) Eight QEs placed in a nanocavity on an `encircled emitter' arrangement and the population dynamics of the (D) central and (E) surrounding  QES, when the system is initialised in the excited field state.
}
\end{figure}
Of course one can have more than just two QEs in a plasmonic nanocavity, and with DNA origami~\cite{Simoncelli2016,Chikkaraddy2018} these can be placed with a spatial precision of $\pm 1$nm. 
Here we choose eight QEs ($N=8$), but keep the discussion generalised as it is applicable for any reasonable number of emitters. 
We initially consider a symmetric QE ring configuration of radius $r$, shown diagrammatically in Figure~\ref{fig:n5}A, where all N-QEs experience the same coupling strength due to the cylindrical symmetry of the plasmon mode. 
The quantum dynamics of this system can be described using only three states: the vacuum state $|0,\vec{g}\rangle$, the excited field state $|1,\vec{g}\rangle$, and an $N$-fold degenerate state with any one of the $N$ emitters excited $|0,\phi\rangle$. Here $|\vec{g}\rangle = \prod^N_i \otimes|g_i\rangle$ is a state in which all of the emitters are in their ground states, and $|\phi_j \rangle = \sigma_+^{(j)} \vert \overrightarrow g \rangle$ is the state where emitter $j$ is excited whilst the rest remain in their ground states. 
The state $|0,\phi_{j}\rangle$ is shown in Figure~\ref{fig:n5}B for this symmetric ring configuration when the system initially in the plasmon excited state; this exhibits a $\sqrt{N}$ enhancement to the effective coupling strength and Rabi oscillation frequency. 
Now, instead consider the emitters placed in an encircled emitter configuration, shown in Figure~\ref{fig:n5}C, with one emitter at the centre of the nanocavity and the other $N-1$ forming a ring of radius $r$. The central and surrounding QEs experience different coupling strentghs and therefore have different quantum dynamics, where $|\phi_1 \rangle$ of the central emitter is no longer degenerate with $|\phi_{j>1} \rangle$ of the ring emitters, shown in Figure~\ref{fig:n5}D and E respectively. 
Analogously to the asymmetric displacement of the two QEs, the central QE remains always strongly coupled to the plasmon, while the surrounding emitters experience slower Rabi oscillations as $r$ increases, until $r \gtrsim 10$nm where they are weakly coupled to the plasmon.

\begin{figure} 
\includegraphics[width=\linewidth]{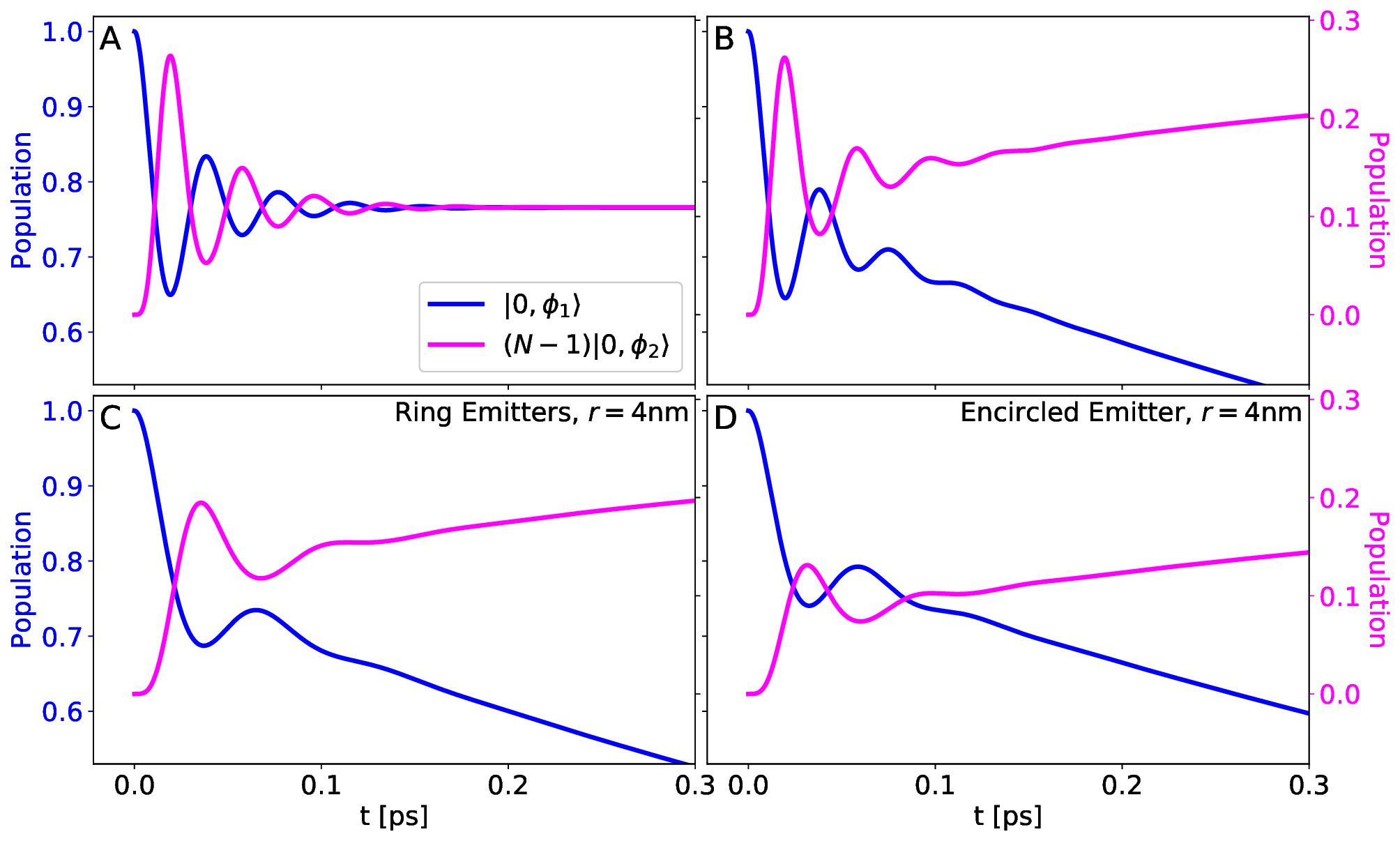}
\caption{\label{fig:n6}The population dynamics for eight QEs in a nanocavity, initialised with one QE (blue) at its excited state and the effective excited state of the other QEs (magenta). The dynamics for the emitters placed at the centre of the cavity, (A) without ($\kappa_{vib}=0$) and (B) with ($\kappa_{vib}=25$meV) QE decoherences, and placing the QEs at a (C) ring and (D) encircled QE arrangements with radius $r=4$nm.}
\end{figure}

We now explore the emergence of the sub-radiant entangled states for a multi-emitter system. 
We initialise the system with one QE in its excited state, when all N-emitters are at the centre of the nanocavity and there is no molecular decoherence ($\kappa_{vib}=0$), Figure~\ref{fig:n6} shows the population of the state $|0, \phi_{1} \rangle$  for the initially excited QE (blue) together with the total population of the other $(N-1)$ QEs (magenta). 
Figure~\ref{fig:n6}B incorporates the decoherence for Cy5  molecules of $\kappa_{vib}=25$meV, and shows the emergence of the sub-radiant entangled states between the N-emitters, that now decays more rapidly because of the higher total QE decoherence. 
If the QEs are placed on a symmetric ring at $r=4$nm, the sub-radiant entangled states are formed, but there are fewer oscillations between the QEs since the coupling strength is much weaker. 
However, the coherent dynamics of the QEs helps maintain their long-lived nature, with the sub-radiant state population matching the same state's population for all QEs at the nanocavity centre. 
For the asymmetric encircled emitter configuration, the system is initialized with the central emitter in an excited state, which maintains the quantum dynamic degeneracy of the $(N-1)$ QEs forming the circle.
However, the varying coupling strength between the central and ring QEs distorts the sub-radiant entangled states. 

Hence, multiple QEs within a plasmonic nanocavity give rise to semi-persistent sub-radiant states at room temperature via the plasmonic dissipation, if the system is initialized appropriately.
These sub-radiant state form within just $\sim 100$fs and persist for $\sim 100$ times longer than the plasmonic excitation.  
They are also formed regardless of the coupling strength value.
The high loss of the plasmonic nanocavity actually enables the emergence of the sub-radiant states, since it causes the super-radiant state to decay very rapidly. 
Their decay depends only on the molecular decoherence, and one can choose molecules, quantum dots or other QEs with more favourable decoherence values than the ones considered here. 
Such systems pave the way towards engineering quantum states at room temperature via dissipation with plasmonic nanocavities, for quantum memory and quantum communication applications, avoiding complex and cumbersome experimental set-ups.

\section{Conclusion}
\label{sec:Conclusion}
We have introduced a theoretical formalism for the integration of quantised plasmons into quantum optical models, with particular attention given to the correct parameterisation of in and out coupling in plasmonic systems. We obtain experimentally measurable quantities from the quantum formalism, with very good agreement between quantum and classically calculated extinction cross-sections. 
Using this method, we demonstrate that multiple emitters in a plasmonic nanocavity form semi-persistent sub-radiant states via the plasmonic disspation, if the system is initialized appropriately. 
The sub-radiant states are independent of the high plasmonic losses and the coupling strength, as long as the QEs are strongly coupled to the plasmon, and survive $\sim 100$ times longer than the plasmonic excitation. 
This work paves the way towards building quantum memories and quantum communication systems operating at room temperature.

\section*{Acknowledgements}
AD gratefully acknowledges support from the Royal Society University Research Fellowship URF\textbackslash R1\textbackslash 180097, Royal Society Research Fellows Enhancement Award RGF \textbackslash EA\textbackslash 181038, Royal Society Research grants RGS \textbackslash R1\textbackslash 211093 and funding from EPSRC for the CDT in Topological Design EP/S02297X/1.

\section*{Supporting Information}
The Supporting Information provides technical details on: (1) Mode volume; (2) Hamiltonian for the driven cavity QED model;  (3) Determination of the in-coupling constant; (4) Quantum extinction cross-section; (5) Single emitter dynamics; (6) Two emitter dynamics and (7) Multi-emitter dynamics.

\bibliographystyle{unsrt}
\bibliography{paper}

\end{document}